\documentclass{aa}

\usepackage{amssymb}
\usepackage{amsmath}
\usepackage{graphicx}
\usepackage{psfig}

\begin{document}

\def\spose#1{\hbox to 0pt{#1\hss}}
\def\lta{\mathrel{\spose{\lower 3pt\hbox{$\mathchar"218$}}
     \raise 2.0pt\hbox{$\mathchar"13C$}}}
\def\gta{\mathrel{\spose{\lower 3pt\hbox{$\mathchar"218$}}
     \raise 2.0pt\hbox{$\mathchar"13E$}}}
\def\Msun{{\rm M}_\odot}
\def\msun{{\rm M}_\odot}
\def\Rsun{{\rm R}_\odot}
\def\Lsun{{\rm L}_\odot}
\def\half{{1\over2}}
\def\RL{R_{\rm L}}
\def\zs{\zeta_{s}}
\def\zR{\zeta_{\rm R}}
\def\dJJ{{\dot J\over J}}
\def\dMM{{\dot M_2\over M_2}}
\def\tKH{t_{\rm KH}}
\def\eck#1{\left\lbrack #1 \right\rbrack}
\def\rund#1{\left( #1 \right)}
\def\wave#1{\left\lbrace #1 \right\rbrace}
\def\dd{{\rm d}}
\def\grs{GRS 1915+105 }

\title{Evidence for local mass accretion rate variations 
       in the disc of GRS~1915+105}

\author{S. Migliari\inst{1,2}
	\and
        T. Belloni\inst{2}}

\offprints{S. Migliari}

\institute{Astronomical Institute ``Anton Pannekoek" 
	University of Amsterdam and Center for High-Energy Astrophysics,\\ 
        Kruislaan 403, NL 1098 SJ Amsterdam, Netherlands e-mail:
        migliari@astro.uva.nl\\   
        \and
        INAF -- Osservatorio Astronomico di Brera,
	Via E. Bianchi 46, I-23807 Merate (LC), Italy e-mail:\\ 
        belloni@merate.mi.astro.it\\
}

\date{Received ???; accepted ???}

\abstract
{We have analysed 16 observations of class $\beta$ of GRS 1915+105 performed
with the Rossi X-ray Timing Explorer. We show the time-resolved evolution
(every 16 seconds) of the X-ray spectral parameters during these observations. The three model independent
states (A, B, C) identified in the color-color diagrams are described here in
terms of the more physical (and model dependent) quantities of the disc and
power law components. As in the color analysis, we observe  specific and
time-asymmetric transitions between states. Within a single
observation, we identified a loop-like cycle of the power law spectral index
that starts and ends with the ending of state C.
There is evidence of local mass accretion rate variations and mass loss during
the long instability intervals (long state C). This is consistent with
simultaneous radio and X-ray observations that show a one-to-one relation
between the long state C and the radio flares, that occur shortly after it:
a loss of disc mass is followed by an ejection of matter in the jet.   
}

\authorrunning{S. Migliari \& T. Belloni}

\titlerunning{Local mass accretion rate variations in GRS 1915+105}

\maketitle


\section{Introduction}

GRS 1915+105 was discovered in 1992 as an X-ray transient by the WATCH all sky
monitor on board GRANAT (Castro-Tirado et al. 1992), and has been bright and
active since then. In 1994, 
radio observations with the VLA (Very Large Array) showed radio knots moving
with apparent superluminal velocities. Mirabel \& 
Rodr\'\i{}guez (1994) interpreted these expansions as collimated relativistic
jet emissions with an inclination of $\sim 70^{\circ}$ from the line of sight. 
GRS 1915+105, located at a distance of $\sim11$~kpc (Fender et al. 1999), 
was the first galactic source to show superluminal expansion. 
A dynamical estimate of the mass of the compact object
based on infrared observations gives M=$14\pm 4$~$\Msun$ and supports the idea
of a black hole (Greiner et al. 2001). 
Since its discovery, GRS 1915+105 was extensively monitored at different
wavelengths. The source shows high variability in the infrared, radio
and X-ray band.
Since 1996, GRS 1915+105 is observed at least weekly by the instruments 
onboard the Rossi X-ray Timing Explorer (RXTE), which confirmed its extreme
X-ray variability even on very short time scale. 
Spectral analysis of selected observations with the
Proportional Counter Array (PCA) suggested that the rapid variability of the
source is caused by a thermal-viscous instability in the inner region of the
accretion disc near the black hole, where the radiation pressure dominates
(Belloni et al. 1997a,b). 
This instability causes the detected inner region of the accretion disc to
disappear.         

Despite the complex variability shown in the 2--30 keV band by the PCA, it
is possible to arrange all the observations from the first 
years of RXTE into 13 distinct classes (Belloni et al. 2000; Klein-Wolt et
al. 2002). This classification is based on light curves, 
color-color diagrams (CD), and color-intensity diagrams, and is
model-independent. 
Furthermore, it is possible to describe the behaviour of all 13 classes using a
combination of only three source states, identified in the CD,
which Belloni et al. (2000) called {\it state A}, {\it state B}
and {\it state C}.  
The 2-30 keV energy spectra of GRS 1915+105 
can be described using the `standard'
model for black hole candidates consisting of the superposition of a
multi-temperature disc blackbody (the soft component, interpreted as an
accretion disc emission) and a power law (the hard component, interpreted as
derived from Comptonization processes).    
These two models are obvious simplifications of the shape of the two 
separate components, but are helpful to characterize the spectral changes of
the source on short time scales.
Adopting a model-dependent approach, we can say that
states A and B correspond to soft energy spectra with an
observable inner region of the accretion disc, 
whereas state C is related to the
`quiescent' state of Belloni et al. (1997a,b), corresponding to an
unobservable inner accretion disc of variable size. 
 
In the radio and infrared energy bands, GRS 1915+105 shows luminosity 
variations both on long and short time scales. Fender et al. (1997)
observed rapid flares in the radio and infrared, and found a clear correlation
between the two. They interpreted these oscillations as synchrotron emissions
from repeated ejection of matter from the system.  
Pooley \& Fender (1997) first suggested a possible association between
radio flares and X-ray dips observed on the same time scale.
Mirabel et al. (1998) reported simultaneous observations in the
X-ray, radio and infrared bands, also inferring a possible relation between the
IR-radio emissions and the long X-ray dip (state C).   
Klein-Wolt et al. (2002) established a
one-to-one relation between the radio flare and the long state C, confirming
the connection of the radio jets and the disc instability in GRS 1915+105.
Ueda et al. (2002) analysed a one week multiwavelength monitoring of
GRS~1915+105, that 
covered a wide energy range from radio to gamma rays. During
`plateau' observations, the radio to infrared spectra are consistent with
coming from an optically thick compact jet, and constrained the infrared
emitting region to a distance of $>10^{3}$~cm from the compact object. In the
X to gamma rays spectra they found a high energy tail with no significant
high energy cutoff untill $\sim 1$~MeV, confirming the results of Zdziarsky et
al. (2001).   
\begin{table}[!h]
\centering
\caption{\small{Observation ID, RXTE orbits, duration of
RXTE orbits and duration of state C (the upper limits indicate that the
observation is interrupted while in state C) of the 16 observations analysed.}}\label{tab_obs}  
\vspace{0.1cm}
\begin{tabular}{l|l|l|l}
\hline
\hline
Obs ID &\# Orb &$\Delta T_{\rm obs}$ (sec) &$\Delta T$ State C (sec)\\ 
\hline
   10408-01-10-00 &a &3480 &432  \\
                &b &3397 &448  \\
\hline
   20186-03-03-01              &a  &1792 \\
                &b  &2197 &704 \\
                &c  &2501 &688 \\
                &d  &2933 &560 \\
                &       &     &592 \\
                &e  &256  &-- \\
\hline
  20402-01-43-00              &a  &9199 &528 \\
                &       &     &560 \\
\hline
 20402-01-44-00               &a &9288 &$>560$ \\
                &      &     &688 \\
                &      &     &672 \\
                &      &     &704 \\
\hline
  20402-01-45-00               &a &2516 &544 \\
                &  &     &528 \\
                &b &2267 &-- \\
                &c &4480 &528 \\
                &d &2025 &--\\
\hline
 20402-01-45-03               &a &3072 &528 \\
                &b &3313 &512 \\
                &c &3425 &720 \\
                &  &     &656 \\
                &d &276  &-- \\

\hline
     20402-01-46-00           &a &383  &-- \\
                &b &3407 &592 \\
                &  &     &624 \\
                &c &3167 &592 \\
                &  &     &640 \\
                &  &     &$>448$ \\
                &d &2797 &528 \\
                &  &     &656 \\
\hline
    20402-01-52-01            &a &2057 &736 \\
                &  &     &$>480$ \\
\hline
     20402-01-52-02  &a &2024 &768 \\

\hline
  20402-01-53-00              &a &8448 &480 \\
                &  &     &464 \\
                &b &3477 &512 \\
\hline
   20402-01-59-00              &a &3367 &544 \\
                &b &3399 &528 \\
                &c &2983 &-- \\
\hline
      30182-01-03-00          &a &1024  &-- \\
                &b &4736  &$>704$ \\
                &c &5248  &-- \\
                &d &11612 &-- \\
\hline
  40115-01-04-00              &a &3254 &896 \\
                &  &     &880 \\
\hline
  40703-01-18-00              &a &8413 &$>496$ \\
                &  &     &992 \\
                &  &     &$>464$ \\
                &  &     &880 \\
\hline
     40703-01-35-00            &a &1779  &688 \\
\hline
    40703-01-35-01            &a &3935 &816 \\
                &b &4096 &$>640$ \\
\end{tabular}
\end{table}

In this paper we focus on the time-resolved evolution of the
X-ray spectral parameters (see also Swank et al. 1998; Markwardt et al. 1999;
Muno et al. 1999) 
during class $\beta$ observations of GRS 1915+105 (after the
classification of Belloni et al. 2000). In this class, all three states
are observed to last long enough to allow the accumulation of reliable
spectral information from RXTE/PCA data.
The paper is structured as follows.
We select (Sect. 2) and analyse (Sect. 3) 16 observations of class $\beta$.
We trace the behaviour and the time evolution of the
spectral parameters derived from fitting the data from A/B/C states
and the relations between them (Sect. 4). We discuss
the results in terms of physical models (Sect. 5). 

\section{Selection of Observations}

As mentioned above, GRS~1915+105 shows complex and rapid variability in
the X-ray band. This has been interpreted as the effect of a thermal-viscous
instability of the inner region of the accretion disc (Belloni et al. 1997a,b;
see also e.g. Taam et al. 1997; Nayakshin et al. 1999; Janiuk et al. 2000;
Zampieri et al. 2000). 
Our aim is to 
follow the temporal evolution of the spectral parameters of the source
during these instability intervals using RXTE/PCA data. 
For the spectral analysis we used {\it Standard2} data, which have a good
spectral resolution for our purposes. However, this choice limits the
classes of observations we can analyse. The {\it Standard2} data have a
timing resolution of 16~s. This implies that in order to trace the evolution
of the spectral parameters during the instability intervals we have to select
observations in which the length of the three states is substantially
longer than 16~s. From the classification by Belloni et
al. (2000), we selected 16 observations of class $\beta$.
The log of the selected observations is shown in
Table~\ref{tab_obs}.   

\section{Data Analysis}

For each observation, we produced light curves at 1~s time resolution using
{\tt Standard1} data in order to identify the long A/B/C intervals and separate
them clearly from faster oscillations.
For each RXTE orbit, we extracted
energy spectra in the range 3--25~keV, with an integration time of 16~s (from
{\tt Standard2} data), We created the
detector response matrices with {\it pcarsp v7.10}, 
and to each spectrum we subtracted the background, estimated 
using {\it pcabackest v.2.1e}.  
In our spectral analysis, we considered the `standard' model for BHCs
which consists in the superposition of two spectral components: a
disc-blackbody and a power law.
Our choice of a simple power law instead of a more accurate Comptonising model
is motivated by the need of a simple model for efficiently performing a large
number of fits with automatic procedures. In order to check that this model
yields meaningful parameters in the framework of Comptonising models, we
fitted a few spectra for each observation with the {\tt bmc} model in
XSPEC. These fits give values of the spectral index $\Gamma$ consistent with
those of a simple power law model, with a small systematic deviation to
larger values of $\sim 3\%$, and comparable distribution of fluxes between
the soft and the hard component.     
It is important to stress that the disc-blackbody model
does not consider relativistic effects and, as we are investigating the inner
region of the disc close to the compact object, this is an approximation that
we have to treat with caution for a more detailed study of the physics
involved. From the normalization of
the thermal component we are able to calculate the inner radius of the
accretion disc ($R_{\rm in}=D\sqrt{(N/cos\theta)}$~km, $D$ is the distance of the
source in units of 10~kpc, $\theta$ is the inclination angle of the disc, and
$N$ is the normalization of the disc-blackbody). However, this
radius is always underestimated, since what we calculate is not the effective
radius of the disc (the radius corresponding to the higher inner
temperature), but the radius derived from the color temperature (see Merloni et
al. 1999).   
All the energy spectra are fitted using this model, modified with
photoelectric absorption (the equivalent hydrogen column density was fixed to
$7\times 10^{22}$~cm$^{-2}$, see Sect. 3.1) and adding a Gaussian emission line
with a fixed central energy which takes into account an excess around 6.4~keV.
We added a 1\% systematic error, the standard estimated residual error on
the calibration. 
The value of the reduced $\chi^{2}$ was usually around 1, although for some
fits it was slightly higher. 
All extractions and fits were performed with automatic procedures.

\begin{table}[!t]
\centering
\caption{\small{Results from the spectral analysis of 15 spectra of class
$\chi$ observations. It is shown the observation ID, the number of the orbit
and the best-fit column density value (N$_{{\rm H}}$).}} \label{tab_chi}
\vspace{0.2cm}
\begin{tabular}{l|l|l}
\hline
\hline
Obs ID  & Orbit & N$_{{\rm H}}$ ($\times 10^{22}$ cm$^{-2}$)\\
\hline
 10408-01-23-00 & a & 7.22$\pm0.05$  \\ 
                & b & 7.28$\pm0.14$  \\ 
                & c & 7.61$\pm0.11$  \\
\hline
 10408-01-24-00 & a & 6.37$\pm0.12$  \\ 
                & b & 6.40$\pm0.13$  \\ 
                & c & 6.72$\pm0.10$  \\ 
\hline
 10408-01-27-00 & a & 7.13$\pm0.13$  \\
                & b & 7.23$\pm0.16$  \\ 
\hline
 20402-01-52-00 & a & 5.76$\pm0.14$  \\
                & b & 5.70$\pm0.12$  \\ 
                & c & 5.78$\pm0.10$  \\
\hline
 30402-01-09-00 & a & 5.78$\pm0.14$  \\ 
                & b & 5.96$\pm0.13$  \\
\hline
 30703-01-16-00 & a & 5.65$\pm0.15$  \\ 
                & b & 5.74$\pm0.16$  \\
\end{tabular}
\end{table}
\begin{figure}[!t]
\centering
\begin{tabular}{c}
\psfig{figure=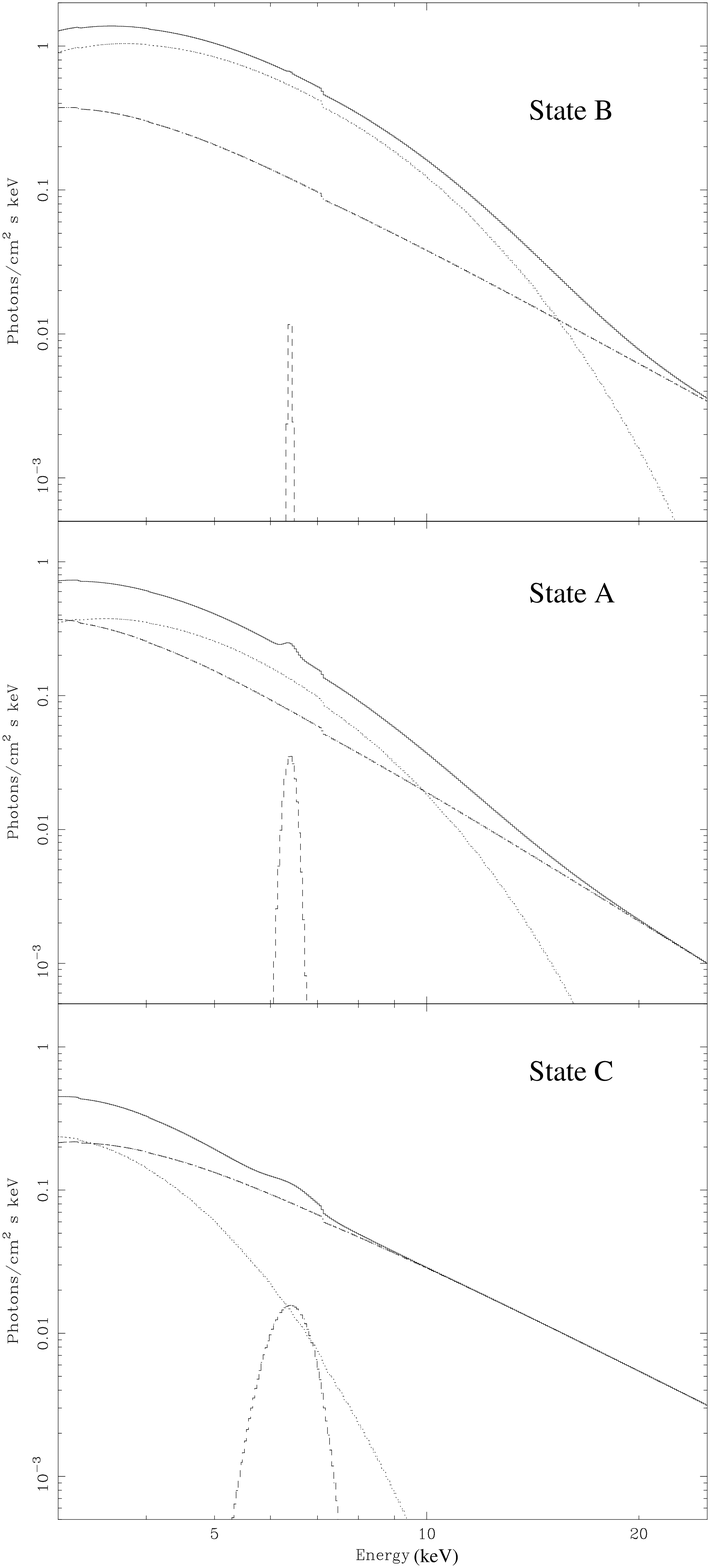,width=8.2cm,height=16cm}
\end{tabular}
  \caption{Examples of fitting spectral models for the three
states A/B/C. The model (solid line) is the superposition of a power law
(dashed-dotted) a disc blackbody (dotted) and a Gaussian emission line
(dashed).}   
  \label{spettri}
\end{figure}

\subsection{Column Density}

Photoelectric absorption modifies the emission spectrum at low energies 
in the PCA band, where the disc component dominates the flux.
The value of the equivalent hydrogen column density
N$_{{\rm H}}$ used to calculate the absorption influences the
normalization of the disc-blackbody component, and thus the value of the
inner radius $R_{\rm in}$. These two quantities (N$_{{\rm H}}$ and $R_{\rm
in}$) are therefore strongly coupled when fitting the spectra. In order to
trace the  variability in the derived inner radius, we had to fix the N$_{{\rm H}}$ value 
(we realistically suppose that the column density does not change within time
scales of the order of half an hour). In order to
find the best N$_{{\rm H}}$ value, we have analysed several class $\chi_{1}$
and $\chi_{3}$ observations (in the classification of Belloni et
al. 2000). The advantage in these classes of observations is that the soft
(disc) component is not detectable in the PCA energy range and therefore 
there are no variations in the soft spectral component that can 
influence the estimated N$_{{\rm H}}$ value. Because of this characteristic and
the absence of significant flux variability over each observation, 
we can accumulate the spectra over a whole RXTE orbit to obtain better
statistics. The observations we have analysed and the best-fit
N$_{{\rm H}}$ values are shown in Table~\ref{tab_chi}. 
For the different observations, the column density varies within the range
$5.8-7.6\times 10^{22}$~cm$^{-2}$, consistent with values found
by other authors (Ebisawa et al. 1995, Belloni et al. 1997b, Muno et
al. 1999). We have analysed the energy spectra of state C intervals for
several class $\beta$ observations of GRS~1915+105, fixing the N$_{{\rm H}}$ at
different values in the range we found for class $\chi$. We found
a general improvement of the $\chi^{2}$ in the estimation of the
spectral parameters using  N$_{{\rm H}}= 7\times 10^{22}$~cm$^{-2}$. The
column density was therefore fixed to $7\times 10^{22}$~cm$^{-2}$ for all
observations discussed below.
We found however that the results remain qualitatively
the same compared to N$_{{\rm H}}=6\times 10^{22}$~cm$^{-2}$.

\begin{figure*}[!t]
\centering
\begin{tabular}{c}
\psfig{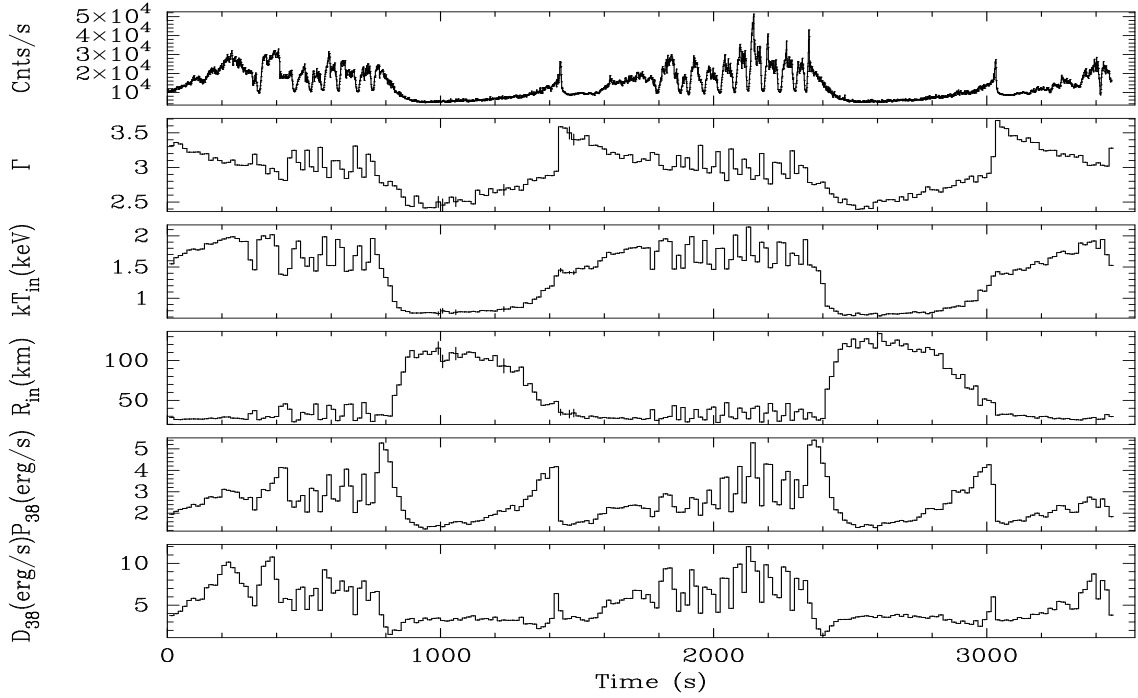}
\end{tabular}
  \caption{Time-resolved evolution of the spectral parameters from the
analysis of an RXTE orbit of the observation 20402-01-44-00.
From top to bottom: 2--60 keV light curve (5 PCUs); 
power law spectral index $\Gamma$; inner disc temperature (kT$_{\rm in}$ in
keV); inner disc radius (R$_{\rm in}$ in km);
3--25~keV luminosity of the power law component {\bf P$_{38}$ in units of
$10^{38}$~erg/s}; bolometric luminosity of the 
disc component {\bf D$_{38}$ (in units of $10^{38}$~erg/s)} calculated using a 
distance of 11~kpc and a disc inclination of $66^{\circ}$ (Fender et
al. 1999). Typical 90\% error bars are shown for a few points.} 
  \label{fig_beta_tutto_44_1all}
\end{figure*}
\begin{figure*}[!t]
\centering
\begin{tabular}{c}
\psfig{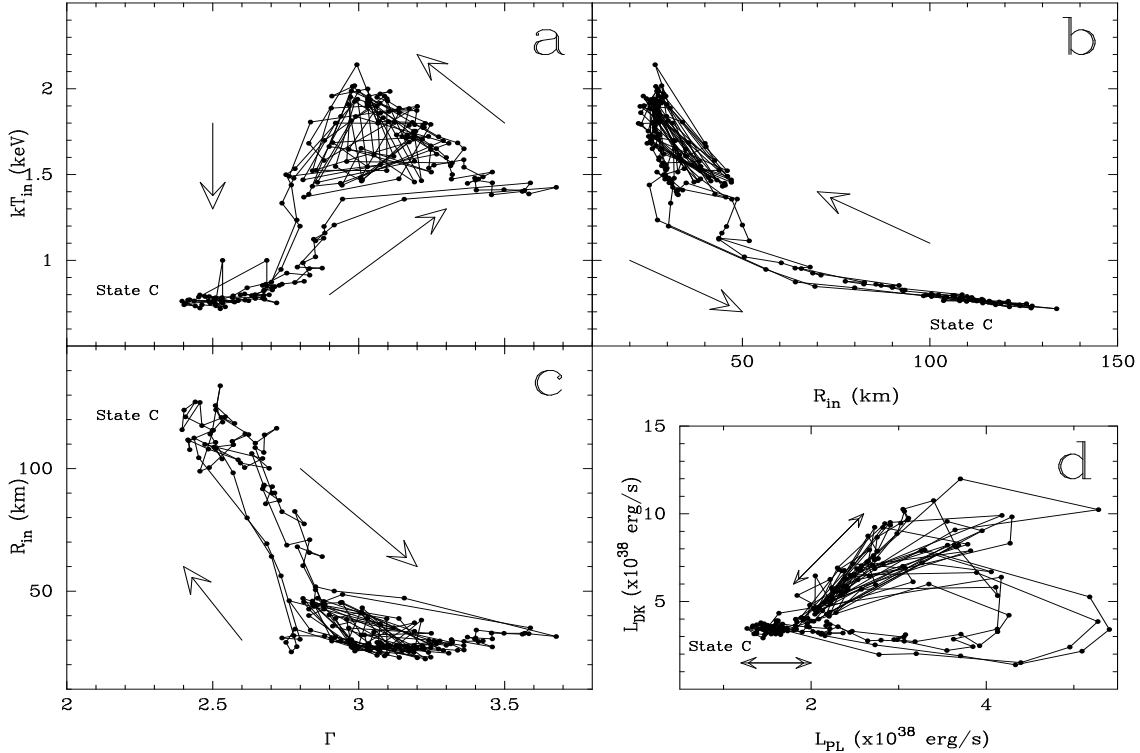}
\end{tabular}
  \caption{Correlations between the spectral parameters of
Fig.~\ref{fig_beta_tutto_44_1all}. Panel {\it a}: power law spectral index 
$\Gamma$ vs inner disc temperature; panel {\it b}: inner
disc temperature vs inner disc radius; panel {\it c}: inner disc radius
vs $\Gamma$; panel {\it d}: luminosity of the disc component vs luminosity
of the power law component.}
  \label{fig_beta_corr_44_1all}
\end{figure*}
\section{Results}

For each RXTE orbit within our observations, we isolated all A/B/C
intervals long enough to obtain significant results using spectra
accumulated over 16 s,
but also analysed the whole orbit (including faster oscillations
for completeness). 
Of course the results from the oscillating parts, 
having been sampled at 16 seconds,
are not covered in an optimal way, but we concentrate our analysis on
the slower variations.
The fitting procedure resulted in very good
values of the reduced-$\chi^{2}$ for almost all the spectra (usually $\sim 1$),
with the exception of
very few spectra, usually during state B. 

In all observations, the light curves during the instability show the same
characteristics: a hard state C, followed by 
an isolated spike at the end of
the state C, a shorter soft low-flux state A shifting slowly into 
state B (see e.g. Swank et al. 1998; Markwardt et al. 1998; Mirabel et
al. 1998; Belloni et al. 2000; Klein-Wolt et al. 2002). In
Fig.~\ref{spettri} we show three representative model spectra for the three
states. Notice the difference in relative contribution of the two main
components.  
As an example, in Fig.~\ref{fig_beta_tutto_44_1all} we show the evolution
of the spectral parameters for the whole first orbit of observation
20402-01-44-00 (for clarity, we show typical 90\%
error bars only for a few points). In the
first panel from the top we show the 2--60~keV light curve, in the
second the power law spectral index, in the third the inner temperature (in
keV) of the accretion disc, in the fourth the inner
radius of the disc (in km), and in the fifth and sixth panels we show the
3--25~keV luminosity of the hard power law component and the bolometric
luminosity of the disc blackbody component as
calculated from the best-fit parameters
(normalisation and $\Gamma$ from the fitting power law, and R$_{\rm in}$ and
T$_{\rm in}$ from the fitting disc blackbody model) using a distance of 11~kpc
and an inclination of the disc of $66^{\circ}$ (Fender et al. 1999) thought to
be perpendicular to the radio jet (see also Maccarone 2002).
The corresponding correlations between parameters are shown in Fig. 2.
Here, the results for the oscillating parts are included, as they give
a general idea of the spectral changes during faster events.

The time evolution of the different parameters can be summarized as follows. 

\begin{itemize}
\item {\it power law parameters:}
From Fig.~\ref{fig_beta_tutto_44_1all}, we can see that 
the evolution of the power law spectral index $\Gamma$ shows a ``cycle''
which starts and ends at the end of a state C interval. Following the first
{\it cycle}, starting $\sim$800 s from the beginning of the observation, 
we notice
that during state C $\Gamma$ follows the light curve, starting from $\sim$2.8
at the beginning, decreasing to 2.4, then increasing again to reach it
maximum ($\sim$3.3) at the end of state C, in correspondence with the ``spike''
in the light curve. Then,
during the state A and B intervals, it gradually decreases until
it reaches a value of about 3.  After this, the faster oscillations start
and $\Gamma$ oscillates in the range 2.8--3.3.
Notice that $\Gamma$ correlates with the count rate during state C,
but not during state B.
The power law flux (fifth panel from the top) shows large variations during
state C, and a much less variable behaviour during states A/B.

\item {\it Disc parameters:}
In the third and fourth panels from the top of Fig.~\ref{fig_beta_tutto_44_1all}, we show the evolution
of the accretion disc parameters (inner temperature and radius). It is evident
the confirmation of the results obtained by Belloni et al. 1997b: during state
C, the inner radius $R_{\rm in}$ increases whereas the inner temperature
$kT_{\rm in}$ decreases, both rather fast. This is
compatible with the idea of a `missing', or simply unobservable, inner
region of the disc. At the end of state C,
the inner radius goes down to a minimum around
30 km corresponding to state A, without a strong jump at the ``spike''.
Moreover, the inner disc radius does not change
much during state A and B, remaining at its minimum value, and even during
the oscillations its variations are rather limited.
The inner disc temperature, on the other hand, remains low $\sim 0.8$ keV
and constant
for much of state C, then towards the end starts increasing. This increase
continues through states A and B up to $\sim$2 keV, again
without dramatic changes in correspondence with the ``spike''.
The ``spike'' is essentially caused by a sharp transition in the properties
of the power law component.
The disc flux evolution  (sixth panel from the top) shows a constant flux
throughout state C, with some evidence of the ``spike'', and an increased
flux during states A/B and during the oscillations. Notice that over the
whole orbit the disc flux dominates over the power law flux.

\end{itemize}
\begin{figure}[!t]
\centering
\begin{tabular}{c}
\psfig{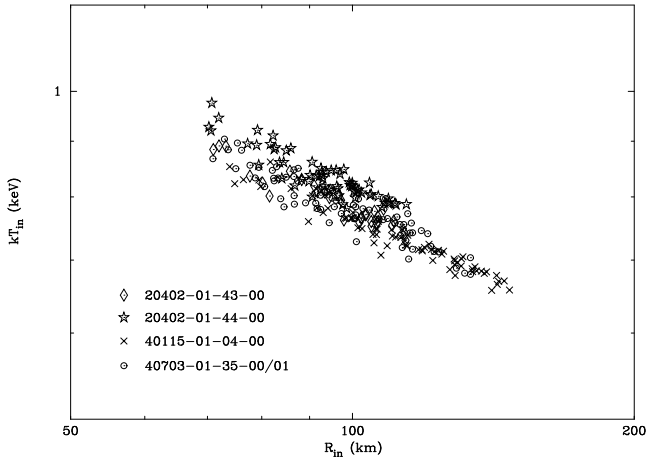}
\end{tabular}
  \caption{Inner disc temperature as a function of inner radius for five
GRS~1915+105 observations. We isolated the intervals of lower count rate
of state C.} 
  \label{fig_beta_r-kt_tuttissimi}
\end{figure}
\begin{figure}[!t]
\centering
\begin{tabular}{c}
\psfig{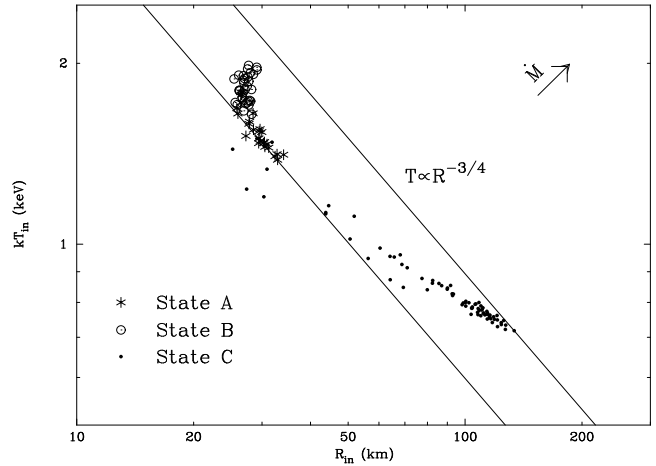}
\end{tabular}
  \caption{Inner disc temperature versus inner disc radius. We plot the points
corresponding to two intervals of instability of the observation
20402-01-44-00: state C, state A and the beginning of state B (without the
rapid flares; see Fig. 2b). We excluded $\sim50$~s of data around the spikes.
}
  \label{fig_beta_r-kt_2_44_1tutti}
\end{figure}
In Fig.~\ref{fig_beta_corr_44_1all}, we show the correlations
between the best-fit parameters: inner disc temperature, inner disc radius
and power law index, plus the disc/power law flux correlation.
Here, obvious and complex correlations are apparent. This behaviour is
common to all observations in our sample. In particular, it is clear that 
the transitions between state C and the softer states is not 
time-reversible. This is evident from the ``loop-like'' shape of the
observed correlations (see also Vilhu \& Nevalainen 1998).

Here we concentrate on the correlation between inner disc radius and
temperature (panels labeled `b'), as this plot gives some indication of the
local accretion rate through the disc at the measured inner radius. 
Despite the uncertainties on the absolute values of radius
measurements mentioned 
above (and caused also by uncertaintes on the distance to the source),
relative values can be useful in determining what happens 
to the disc during the instability intervals.   
From Fig.~\ref{fig_beta_corr_44_1all}b, one can see that during state C there is a clear inverse
correlation between inner radius and temperature, correlation which is 
rather similar throughout the whole of state C. However, while entering
and leaving state C, the source follows rather different paths, mostly
due to the presence of the ``spike'' at the end of state C. As usual, 
faster oscillations are not optimally sampled and result in a noisy loop
structure. 
Vilhu \& Nevalainen (1998) found a ring-shaped pattern in the CD during fast
quasi-regular variations in the light curve count rate (class $\rho$
observations), where the source is rapidly oscillating between state C and
state B. We notice that a loop (ring-like) structure can be seen also in
Fig.~\ref{fig_beta_corr_44_1all}d (power law luminosity  L$_{\rm pl}$ vs disk blackbody luminosity L$_{\rm dk}$),
where the transitions between the hard state C and the soft states A/B have a
ring shaped pattern. The difference from the fast variations seen in
class $\rho$ observations seems to be only the duration of the single
states. In $\beta$ observations the source lies for a longer time in each
state (A/B/C) where the power law and the disk blackbody luminosities form two
elongated branches. Faster ring-like transitions happen between these two branches. This suggests that the physical processes that cause state transitions
are the same independently on the time that the source spends in one state
(seconds as in class $\rho$ or tens of minutes as in class $\beta$).  

\section{Discussion}

The evolution in the R$_{\rm in}$--$kT$ plane can be read in terms of 
variations in the local mass accretion rate flowing through the inner
radius of the optically thick part of the accretion flow. 
In Fig.~\ref{fig_beta_r-kt_tuttissimi}, we have isolated the
interval corresponding to the lower count rate of state C for five
observations.  
We notice that all the points of the different observations follow the
same correlation, that can be fitted with a power law with slope $\sim
-0.4$. At constant accretion rate, from a standard Shakura \& Sunyaev (1974)
disc, a slope $-$0.75 would be expected.  
This indicates that there are systematic variations in the local disc mass
accretion rate during state C, where ${\dot {\rm M}}$ is at a relative maximum
at the bottom of state C, when the inner disc radius is larger. In order to
follow the parameter evolution throughout states A and B,
in Fig.~\ref{fig_beta_r-kt_2_44_1tutti} we plot the points from two 
instability intervals of the observation 20402-01-44-00. Here two lines
mark the relation 
$T_{\rm in} \propto R_{\rm in}^{-3/4}$ for two values of the accretion rate
(increasing to the upper right). The correlation shown in Fig. 5 is evident
in the bottom part of the plot. However, state A points, when the inner
radius is around 35~km, follow precisely the expected $-$0.75 slope,
indicating that the local accretion rate remains constant during state
A. Then, when state B is reached, the inner radius reaches its minimum 
value, probably corresponding to the innermost stable orbit around the black
hole, while the temperature continues to grow, leading to a further
increase of the disc accretion rate.

This overall behaviour indicates that there is a loss of matter in the inner
region of the disc during the instability (state C), with the local 
disc accretion rate oscillating between two values during the instability (the
difference in local mass accretion rate between these two values is of a few
$10^{-8}$~$\Msun$/yr).     
This is consistent with the results by Klein-Wolt et al. (2002), who find a
one-to-one relation between long state C intervals and radio flares, 
and that these outflow emissions occur shortly after the instability 
intervals, where we found a loss of matter in the disc.

Finally, we notice that the minimum inner radius of $\sim$30 km reached 
during state B appears to be the same in all observations in our sample, 
consistent with its association with the minimum stable orbit around the 
black hole.

\section{Conclusions}

We have analysed 16 class $\beta$ PCA/RXTE observations of GRS~1915+105, and
studied the time-resolved evolution of the spectral parameters.  
We can summarise our results as follows:

\begin{itemize}
\item[$\bullet$] The three states (A, B, C) identified in the CD (Belloni et
al. 2000) are described here in terms of the more 
physical quantities of the disc and power law components.  
As in the color analysis, we observe specific and time-asymmetric 
transitions between states. These transitions, already
noticed in Belloni et al. (2000), are now translated into more physical
(and model dependent) parameter evolutions. A large degree of
`clustering' of spectral and timing parameters in a few states were already
noticed by Markwardt et al. (1998) studying one class $\beta$ observation.
A loop-like structure is found and a ring-shaped pattern as found by Vilhu \&
Nevalainen (1998) in (fast flux oscillating) class $\rho$ observations can be
seen also in $\beta$ observations (Fig.~\ref{fig_beta_corr_44_1all}d). The only
difference with the class $\rho$ seems to be the presence of two `branches'
that are formed because of the longer duration of the hard (C) and the soft
(A/B) states.    

\item[$\bullet$] Within a single observation, the power law spectral index
$\Gamma$ has a {\it cycle} that starts and ends with the ending of state C
(i.e. in correspondence with the spike).

\item[$\bullet$] There is evidence, consistent with simultaneous radio and
X-ray observations (see Klein-Wolt et al. 2002), of local mass accretion rate
variations and mass loss during the long instability intervals (long state
C).

\end{itemize}

\noindent
{\em Acknowledgements.} TB thanks the Cariplo Foundation for financial
support. We would like to thank Rob Fender for comments on a draft of this
manuscript.

\end{document}